\documentclass[journal]{IEEEtran}
\usepackage{cite}

  \usepackage[pdftex]{graphicx}
  \graphicspath{{figures/}}
\usepackage{amsmath}
\usepackage{amsfonts,bm}

\def\rvepsilon{{\mathbf{\epsilon}}}
\def\rvtheta{{\mathbf{\theta}}}

\def\rvn{{\mathbf{n}}}

\def\rvx{{\mathbf{x}}}
\def\rvy{{\mathbf{y}}}

\def\rvepsilon{{\mathbf{\epsilon}}}
\newcommand{\E}{\mathbb{E}}
\def\rvtheta{{\mathbf{\theta}}}

\begin{document}
%
\title{Photon-counting CT using a  Conditional Diffusion Model for Super-resolution and Texture-preservation}
%
%
%

\author{Christopher Wiedeman$^{1 \dagger}$,
       Chuang Niu$^{1 \dagger}$,
       Mengzhou Li$^{1}$,
       Bruno De Man$^{2}$,
       Jonathan S Maltz$^{3}$*,
       and Ge Wang$^{1}$*
\thanks{$^1$: Department of Biomedical Engineering, Center for Biotechnology \& Interdisciplinary Studies, Rensselaer Polytechnic Institute, Troy, NY USA}
\thanks{$^2$: HealthCare Technology and Innovation Center, GE HealthCare, Niskayuna, NY}
\thanks{$^3$: Molecular Imaging and Computed Tomography, GE HealthCare, Waukesha, WI, USA}
\thanks{$\dagger$: Equal Contribution}
\thanks{*: Corresponding Authors}
}

\maketitle
\begin{abstract}
Ultra-high resolution images are desirable in photon counting CT (PCCT), but resolution is physically limited by interactions such as charge sharing. Deep learning is a possible method for super-resolution (SR), but sourcing paired training data that adequately models the target task is difficult. Additionally, SR algorithms can distort noise texture, which is an important in many clinical diagnostic scenarios. Here, we train conditional denoising diffusion probabilistic models (DDPMs) for PCCT super-resolution, with the objective to retain textural characteristics of local noise. PCCT simulation methods are used to synthesize realistic resolution degradation. To preserve noise texture, we explore decoupling the noise and signal image inputs and outputs via deep denoisers, explicitly mapping to each during the SR process. Our experimental results indicate that our DDPM trained on simulated data can improve sharpness in real PCCT images. Additionally, the disentanglement of noise from the original image allows our model more faithfully preserve noise texture.  
\end{abstract}

%

\section{Introduction}
\IEEEPARstart{X}{-ray} CT is a preferred modality when high-resolution 3D tomography of internal anatomy is required. Improving CT resolution beyond its current capability is crucial for tasks such as coronary plaque and inner-ear imaging. Unfortunately, x-ray resolution is fundamentally limited in conventional energy-integrating detectors (EIDs), which must convert x-rays into light directed toward photo-diodes. Each pixel must be encased with reflective borders to prevent optical crosstalk, and reducing pixel size increases the relative area of  ``dead space'' between pixels, reducing dose efficiency.

Novel photon-counting detectors (PCDs) promise a multitude of improvements over EIDs for CT imaging. PCDs convert x-ray photons directly into charge clouds, which are registered by electrodes as discrete events rather than as aggregate energy. This more direct process can increase quantum efficiency. In addition, the counting of discrete photons minimizes electronic noise effects, and can produce spectral images using a single x-ray exposure. More importantly, pixels can be spaced much more densely without compromising dose efficiency, as there are no reflective septa. Photon-counting CT (PCCT) has demonstrated improved resolution and image quality over EID systems \cite{willemink_photon-counting_2018}.

Despite these advantages, PCCT is subject to other resolution-limiting considerations. First, the focal spot of the x-ray source can be a limiting factor, especially when imaging larger anatomies, where higher tube current density typically necessitates the use of larger focal spots. Secondly, as pixel density increases, the likelihood of charge sharing, in which the charge associated with a single incoming photon is spread across multiple neighboring pixels, also increases. Thirdly, quantum interactions within the detector material distort spectral and spatial information. For CdTe/CZT (high Z) detectors, this distortion is caused by K-escape fluorescence depositing secondary $\approx$30~keV photons at some distance from the point of primary interaction. For edge-on deep silicon detectors (low Z), Compton scatter becomes a concern. Improvement in hardware and fabrication cannot easily correct these drawbacks, as they occur at the atomic level. As such, researchers have turned to algorithmic methods for further resolution improvement, but  directly reversing the complex processes listed above with physics models is computationally impractical. Thus, we investigate the deep learning (DL) algorithms -- specifically diffusion generative models -- for enhancing PCCT resolution in the image domain.

DL image super resolution (SR) and deblurring has achieved broad success in the computer vision research. Architecturally, most deep SR models are convolutional neural networks, where significant improvements have come from optimizing the modular sequence (EDSR) and residual connections (RCAN) within networks \cite{lim2017enhanced, zhang2018image}. SR enhancers have also been investigated in a plug-and-play fashion, where a deep SR module is used a prior in place of a denoiser \cite{zhang2019deep}.  Generative adversarial networks (GANs) have provided a superior loss objective for producing perceptually realistic high-fidelity images \cite{ledig2017photo}. PULSE has produced better SR results by learning a high-resolution data manifold and searching for images that properly downscale to the input \cite{menon2020pulse}.

More recently, denoising diffusion probabilistic models (DDPMs) have demonstrated and largely defined state-of-the-art performance in a multitude of generative and inverse problems \cite{ho2020denoising, song2022solving, chung2022improving, xia2022patch}.  DDPMs are models trained to gradually reverse a Wiener process performed on target samples over many timesteps, eventually learning to recover new samples from pure Gaussian noise. Compared with GANs, DDPMs produce better diversity as they do not suffer from mode collapse, and have achieved superior image super resolution \cite{dhariwal2021diffusion, saharia2022image}. However, model convergence and lengthy inference times due to the iterative denoising process restrict the use of DDPMs in high-dimensional problems.

Although desirable, translating SR to the medical image domain is complicated by several obstacles. 3D medical images can have millions of voxels, and scaling both the training data and model is computationally challenging, particularly for the aforementioned DDPM. Furthermore, these DL methods require either paired or unpaired low resolution (LR) and high resolution (HR) data. To produce LR datasets, downsampling and linear shift-invariant blurring kernels are often applied to degrade existing images. A model learns to recover the original images from LR-HR pairs \cite{yu2017computed, you2019ct, jiang2020ct}. However, models trained with this approach frequently fail to perform satisfactorily on real images, most likely because the degradation operator inadequately models the true mapping between low and high resolution. This is especially true for PCCT, where the complex interactions that influence resolution and noise cannot be easily captured with simple operators \cite{li2020x}. Finally, it is crucial to consider and preserve noise texture while enhancing features in CT, as this noise is a key feature for trained readers. As such, for clinical use, SR models must be robust to noise level and avoid suppressing or distorting noise. For example, denoising may suppress diagnostically-relevant noise textures, such as those that identify interstitial lung disease.

The following work investigates the use of DDPMs for PCCT SR, and is a continuation of \cite{niu_fully3D}. To synthesize paired data, CatSim is used to generate realistic, low-resolution counterparts to real PCCT head and chest scans, accounting for pixel cross-talk and noise. The degradation is modulated by altering the size of the digital phantom, focal spot and detector pitch. During testing, the trained model is applied to the original PCCT images to observe whether the learned features transfer to the target image domain. Interestingly we found that training a conditional DDPM for SR leads to oversmoothing of CT noise in homogeneous regions. As such, we explore two new input/output schemes, where deep denoisers are first applied to disentangle noise from the signal. Feeding noise and signal separately is used to emphasize proper noise mapping during SR.

\section{Methods}

\subsection{Data and Simulation}
Twenty chest and seven head PCCT scans from GE HealthCare's Deep Silicon prototype PCCT scanner were obtained as digital image phantoms. The CatSim PCCT module was used to simulate two instances of noised, low resolution (LR) and noised, high resolution (HR) counterparts to the phantoms \cite{deman2007catsim, wu2022xcist}. The two LR noise instances are used to train a LR denoiser using Noise2Sim \cite{niu_noise2sim_2020}. The noised HR images paired with the original phantoms are used to train a supervised HR denoiser. One chest and one head scan and their corresponding simulated images are withheld from training for validation and testing. 

All scans are simulated with 120~kVp tube potential and 1~s rotation period, with geometry and hardware specific to the GE system system. Exact parameters for focal spot, pixel size, and detector pitch are not disclosed for proprietary reasons. Noisy HR images are simulated with a 350~mA source current, as well as a 50\%-reduced focal spot size, increased number of views, and no pixel cross-talk, to ensure noise addition without resolution loss. LR images are simulated with cross-talk modeled, and a 220~mA tube current. As simulating with system-specific parameters results in minimal resolution loss, phantoms are digitally shrunk by 50\% in all directions, effectively increasing the degradation during scanning. The assumption is that learning to resolve shrunken features under a realistic forward model will transfer to resolving smaller features in real images. All volumes are reconstructed using filtered back projection, with voxel sizes and locations identical to the original phantoms.

\subsection{Conditional DDPM and Noise-Preservation}
\subsubsection{Conditional DDPM}
Following \cite{niu_fully3D}, we formulate PCCT SR as a conditional generation problem.
Given a Gaussian noise image $\rvx$ and a conditional LR PCCT image $\rvn$, we aim to predict the HR PCCT image $\rvy$ through learning the conditional distribution $p(\rvy | \, \rvx)$.
Here $p(\rvy | \, \rvx)$ is approximated with a reverse Markovian diffusion process, where each iteration step is parameterized with the neural network $f_{\rvtheta}$.
For optimizing $f_{\rvtheta}$, DDPM curated a forward Markovian process with fixed steps.
Specifically, the forward process gradually adds Gaussian noise to an HR image, resulting in an image series $\rvy_0 \rightarrow \rvy_1 \rightarrow \cdots \rightarrow \rvy_T$, where the noise level gradually increases with time step $t$, $\rvy_0$ and $\rvy_T$ are the HR and pure Gaussian noise image respectively, and $T$ is the predefined number of iteration steps. The training loss can be derived as:
\begin{equation}
\label{eq_loss}
    L = \E_{\rvx, \rvy} \, \E_{\rvepsilon, \gamma} \, || f_{\rvtheta}(\rvx, \sqrt{\gamma} \, \rvy_0 + \sqrt{1-\gamma} \, \rvepsilon, \gamma) - \rvepsilon ||_l^l,
\end{equation}
where the loss function is to predict the Gaussian noise added in $\rvy_t$, conditioned on the LR image and the noise level $\gamma$, we set $l=1$, $(\rvx, \rvy)$ is a paired training sample, $\gamma \sim p(\gamma)$ is defined as in \cite{saharia2022image}, and $\rvepsilon$ is normally-distributed noise.
The inference of reverse Markovian diffusion for generating HR PCCT image is:
\begin{equation}
\label{eq_inf}
    \rvy_{t-1} \leftarrow \frac{1}{\sqrt{\alpha_t}} \Big[\rvy_t - \frac{1-\alpha_t}{\sqrt{1-\gamma_t}} f_\rvtheta(\rvx, \rvy_t, \gamma_t)\Big] + \sqrt{1 - \alpha_t}\, \rvepsilon_t.
\end{equation}
Thus, the HR image ($t=0$) can be iteratively generated from a pure Gaussian noise image ($t=T$) conditioned on an LR image.
\subsubsection{Noise-Preservation Schemes}
According to the Noise2Sim theorem~\cite{niu_noise2sim_2020}, neural networks cannot be trained to predict random noise from independent noisy inputs such that noise is suppressed while signals are preserved. 
In practice, we find that conditional DDPM also suppresses some noise components in HR images when conditioned on LR images with independent noise, resulting in inconsistent noise patterns (Figure~\ref{fig:psd_plot}).
To address this problem, we propose disentangling the noise components from the noisy LR input and thus enable the DDPM to generate an HR image more conditioned on LR contents. 
We explore two alternative schemes for the conditional DDPM: (1) \emph{DDPM-noise-split}: stack each denoised LR image and the disentangle noise as a two-channel conditional input; (2) \emph{DDPM-denoise}: use the denoised LR image only as the conditional input. Both are compared to \emph{DDPM}, which is fed the regular LR input.
In this study, we aim to not only develop SR models with simulation data but also apply SR models to real PCCT data.
Therefore, we use the Noise2Sim method to denoise both simulated LR and real PCCT images for the proposed schemes.
Specifically, we use two simulated LR images with independent noise as inputs and targets for training an LR deep denoiser.
To train the real PCCT denoiser, we simulate HR images with higher noise as input, to predict the real PCCT image.
The denoising results shown in Figure~\ref{fig:noise_img} demonstrate the effectiveness of the trained deep denoisers.

\subsubsection{Implementation Details}

In our implementation, 128 $\times$ 128 patches are randomly cropped for training the networks. We use the same network architecture as in \cite{niu_fully3D}, and the attention is applied to the layer with the smallest spatial dimension. The batch size is 4 and the Adam optimizer is used with a learning rate of $10^{-4}$. The number of sampling steps for DDPM is set to 2,000, the number of training iterations to 500,000, and all other hyper-parameters are set equal to those employed in \cite{niu_fully3D}.
We use the same denoising network architecture (two-channel UNet) as in \cite{niu_noise2sim_2020}.
We train both denoising networks for 10,000 iterations using the Adam optimizer with a batch size of 8 and a learning rate of $10^{-5}$.
Our implementation is based on PyTorch. While it is well-known that automatic mixed precision improves training speed, we find that this sometimes results in unstable training in our experiments; this technique is not used in this study.

\section{Results and Discussions}
\begin{figure}[!t]
\centering
\includegraphics[width=3 in]{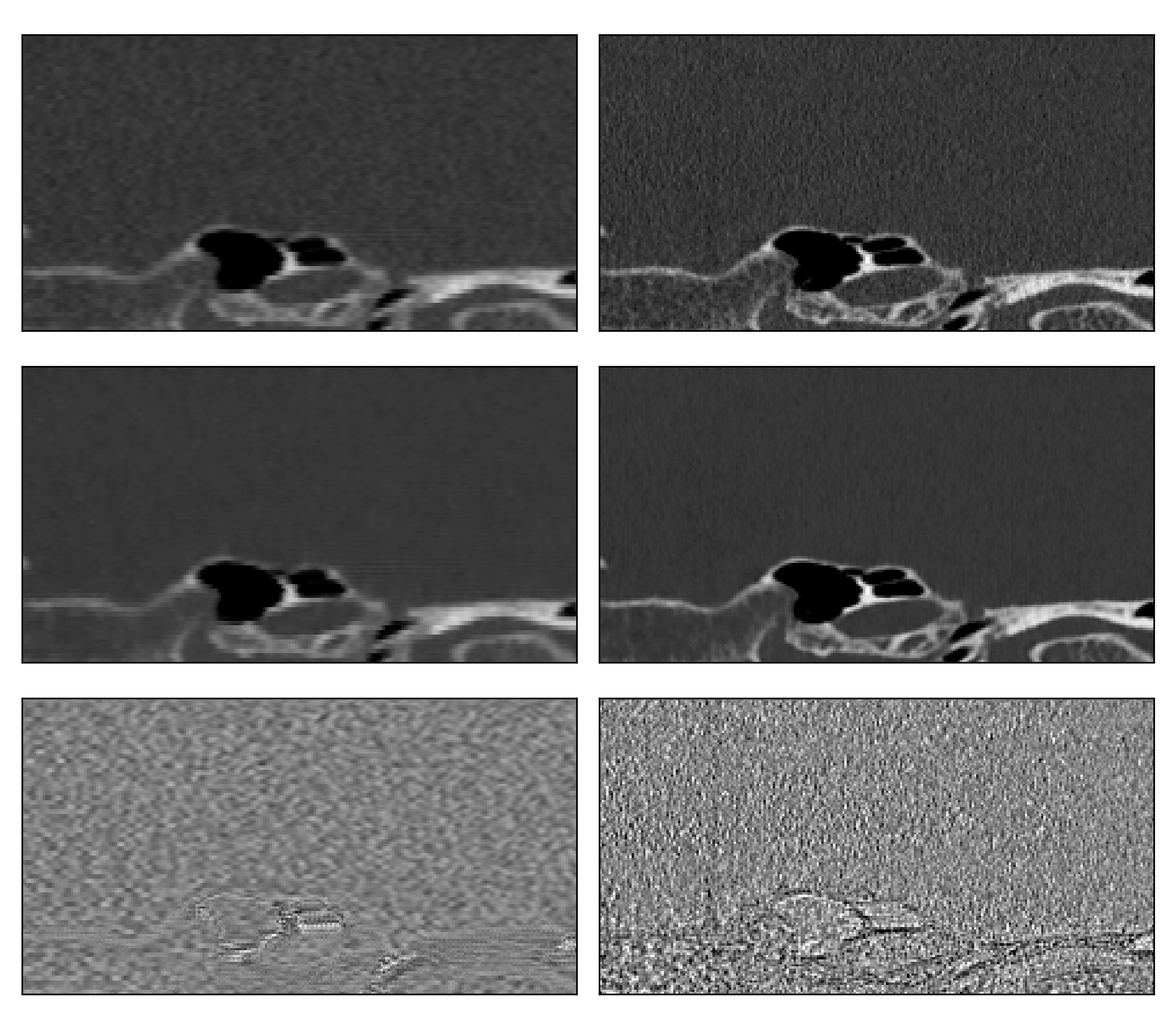}
\caption{Denoising example results for a simulated low-resolution (left column) and true PCCT (right column) patch. Top row: original image, [-500, 2000]~HU. Middle: Denoised image [-500, 2000]~HU. Bottom: Noise map, [-200, 200]~HU.}
\label{fig:noise_img}
\end{figure}

We examine temporal bone images in the coronal plane. Figure \ref{fig:noise_img} illustrates an example of both the LR and HR deep denoisers. Overall, it appears that both denoisers successfully separate noise from signal. Importantly, although the HR denoiser was trained to map from noisier simulations to original images, the model successfully denoises the original PCCT images. Both denoisers still inevitably remove some high frequency signal, which can be seen from the residual bone structures in the noise image. However, this fraction of information is retained during the SR enhancement for \emph{DDPM-noise-split}, as it is explicitly fed both signal and noise channels. 

\begin{figure}[!t]
\centering
\includegraphics[width=3 in]{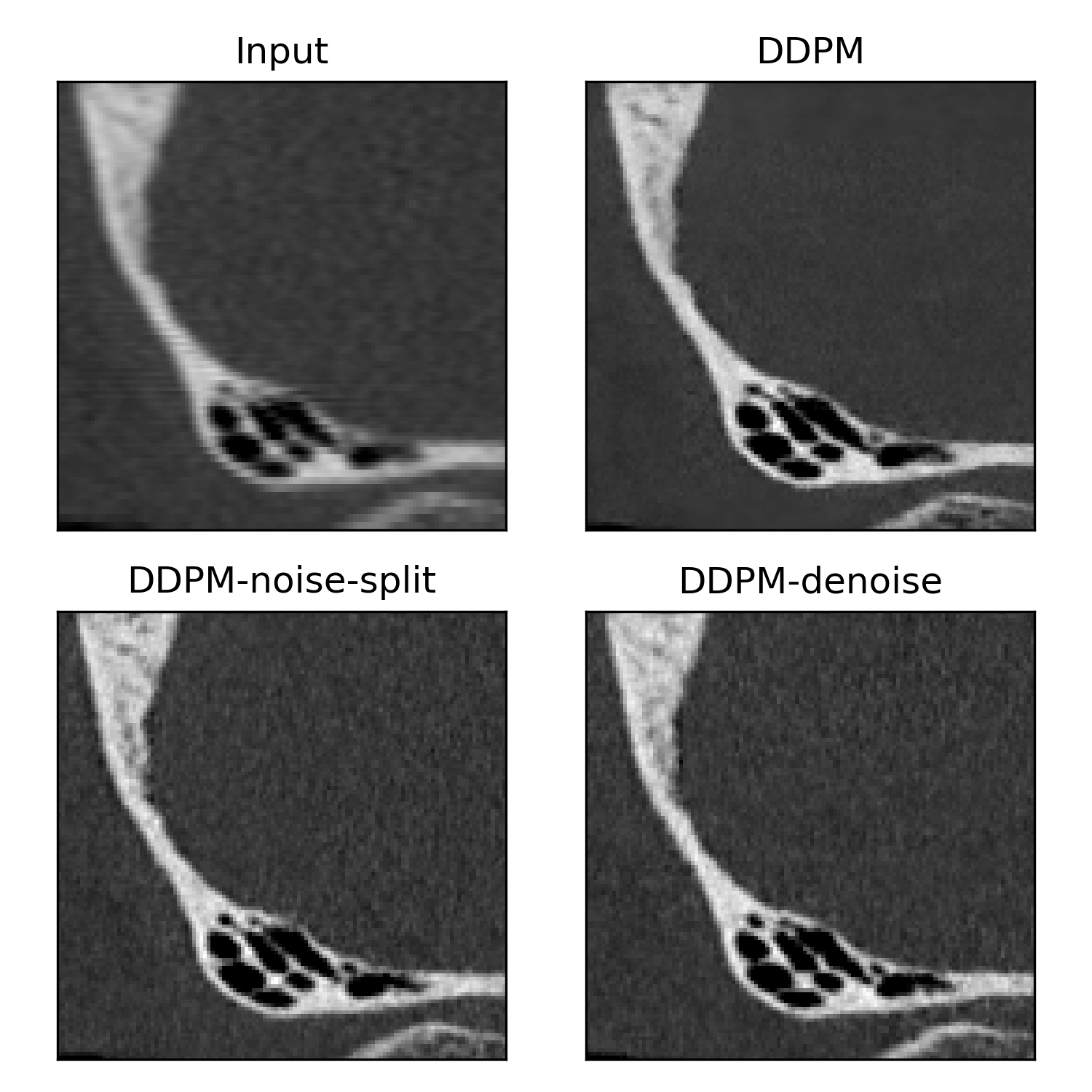}
\caption{Qualitative SR output comparison for simulated validation image. Input is the simulated LR image. [-500, 2000]~HU.}
\label{fig:lowres_imgs}
\end{figure}

\begin{figure}[!t]
\centering
\includegraphics[width=3 in]{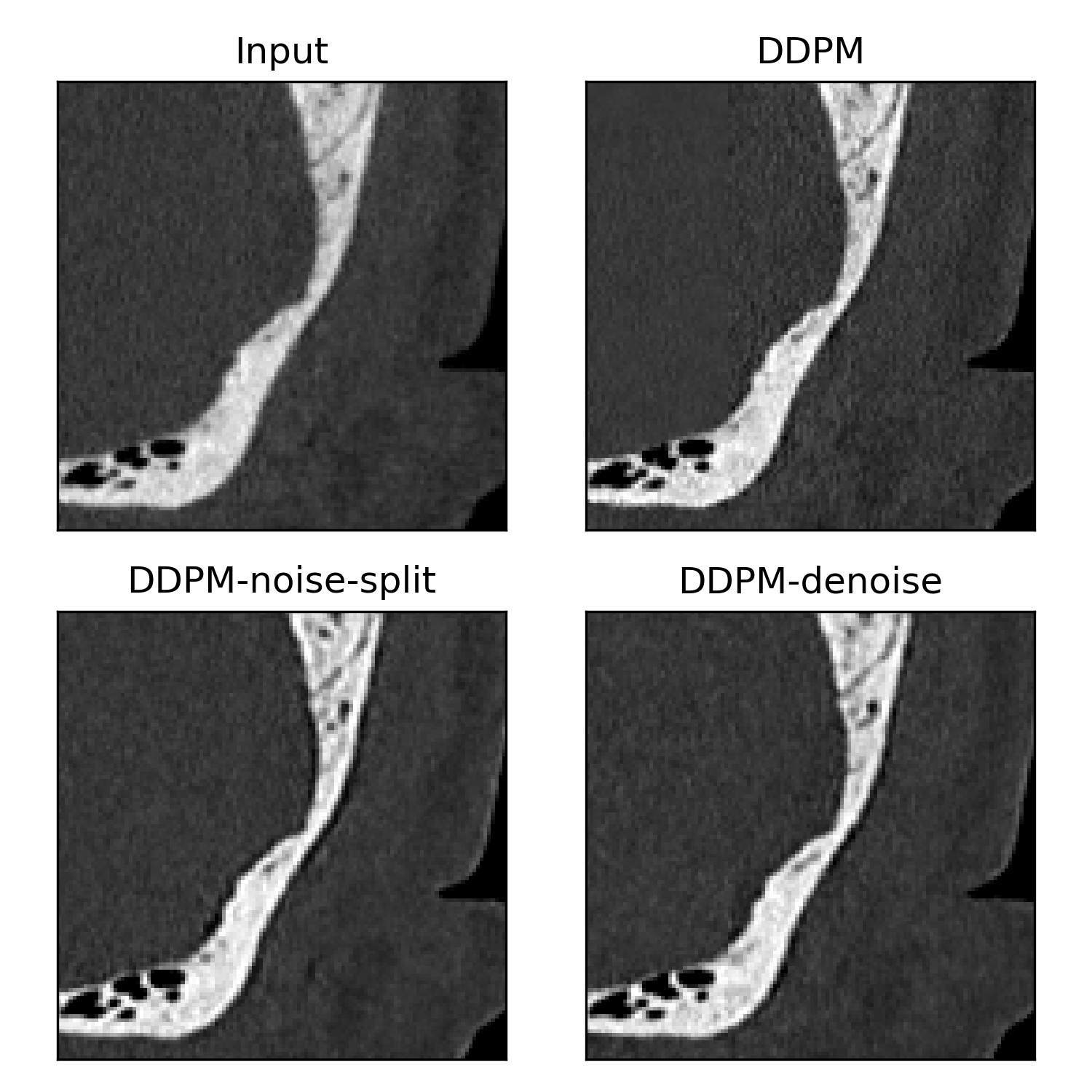}
\caption{Qualitative SR output comparison for real PCCT test image. Input is a real PCCT image. [-500, 2000]~HU.}
\label{fig:highres_imgs}
\end{figure}

Figures \ref{fig:lowres_imgs} and \ref{fig:highres_imgs} compare the SR results for all models on temporal bone regions in the simulated validation and test scans, respectively. We can see that all DDPMs substantially sharpen bone features relative to the simulated LR image. The learned feature sharpening also appears to transfer to the real PCCT scans. However, this enhancement is understandably more modest. In both validation and test cases, \emph{DDPM} undesirably suppresses noise texture, especially in the soft tissue regions. In contrast, \emph{DDPM-noise-split} and \emph{DDPM-denoise} better preserve this texture, indicating that these schemes can more effectively map to the desired noise.

\begin{figure}[!t]
\centering
\includegraphics[width=3 in]{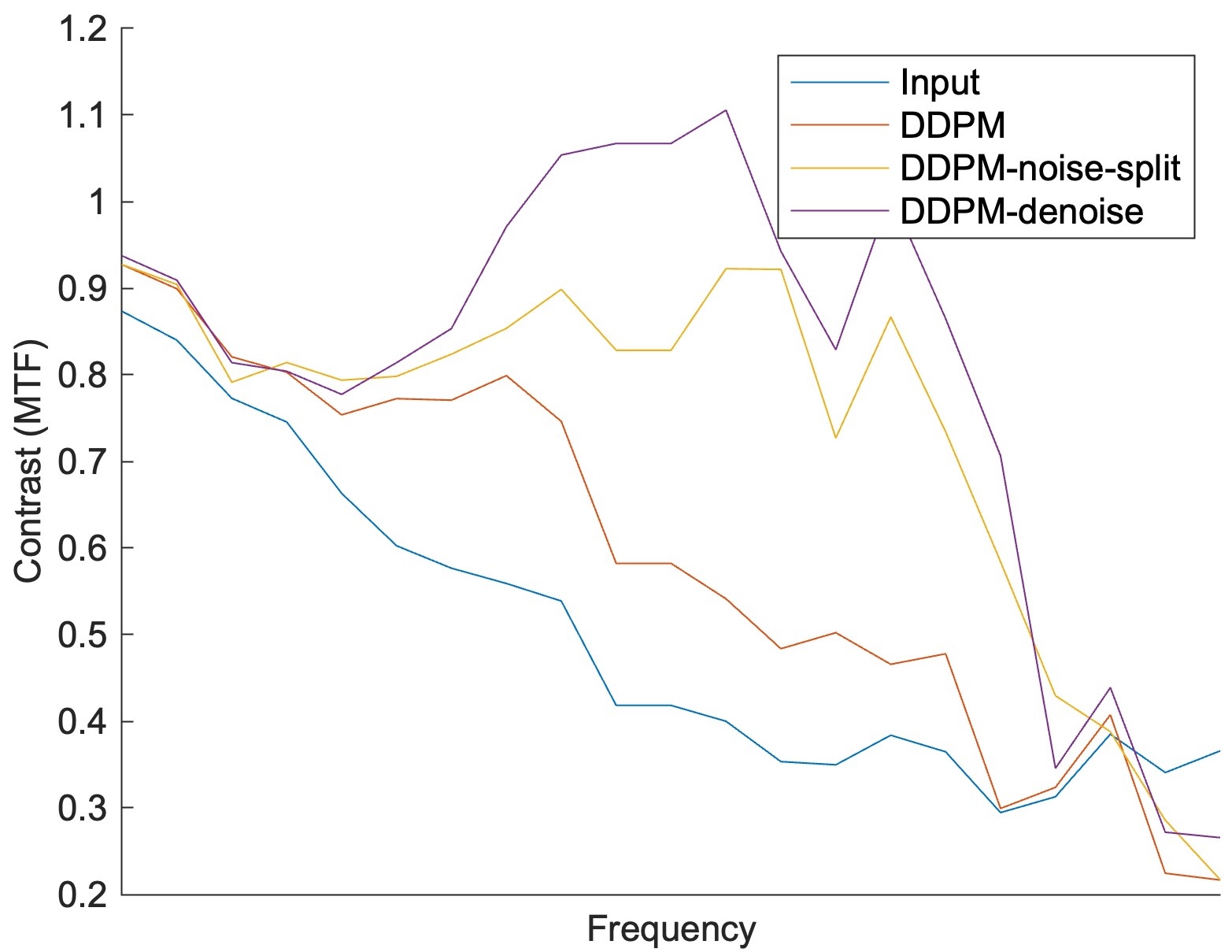}
\caption{MTF for the input compared with, SR models calculated from an actual PCCT scan of CatPhan714. Frequency units are redacted for proprietary reasons.}
\label{fig:mtf}
\end{figure}

Quantitatively, we apply each model to an image slice a GE-PCCT scan of CatPhan714 and calculate the modulation transfer function (MTF), which is plotted in Figure \ref{fig:mtf}. All three DDPMs enhance the MTF, although \emph{DDPM-noise-split} and \emph{DDPM-denoise} interestingly achieve scores < 1 at some frequencies, which can occur when the model boosts contrast above the desired target. It also should be noted that the image slice was in the axial plane while the models have only been trained on coronal plane images. This domain mismatch will be corrected in the future as we train SR models in other planes. 

\begin{figure}[!t]
\centering
\includegraphics[width=3.7 in]{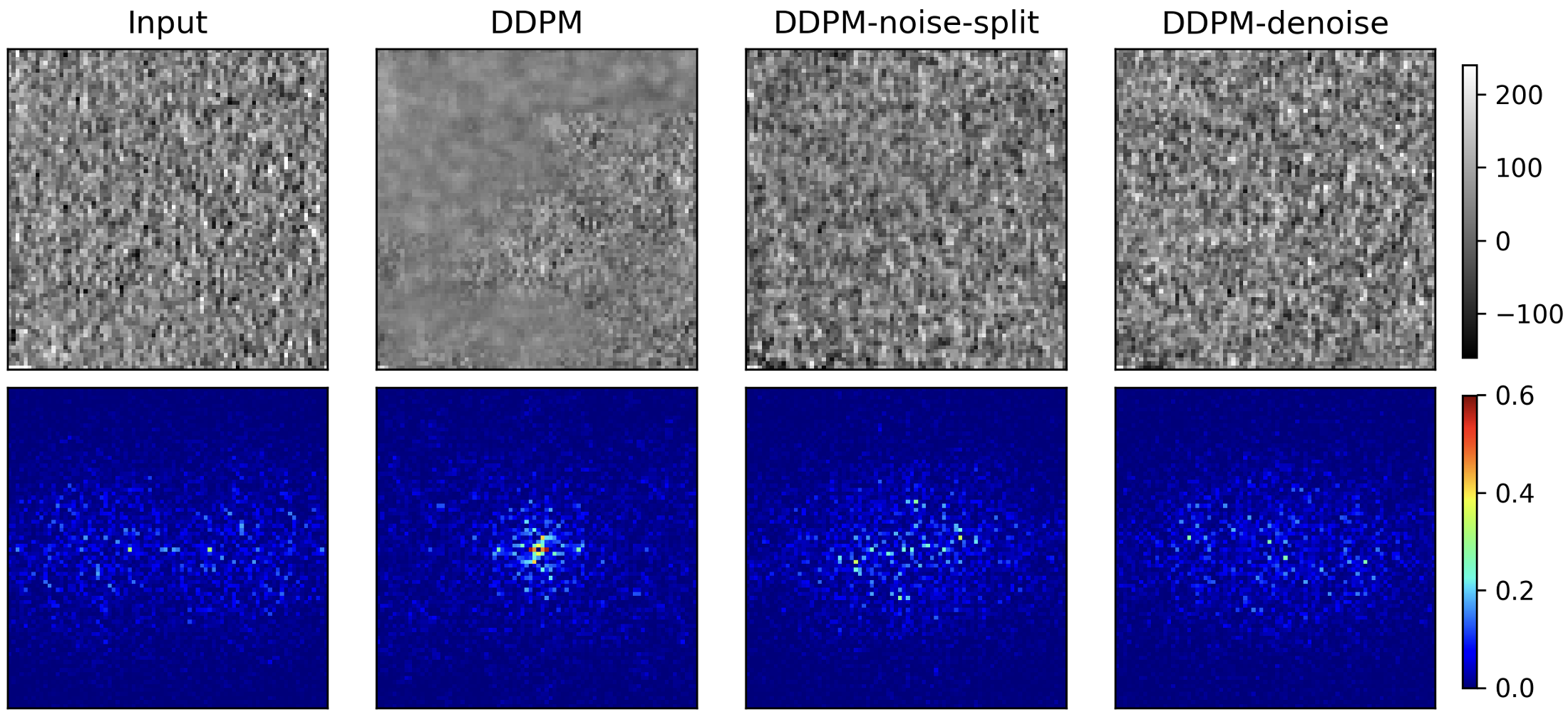}
\caption{Top row: Example patch from homogenous soft tissue region in real PCCT image (HU). Bottom row, power spectral density of patch noise (\% total power).}
\label{fig:psd_plot}
\end{figure}

As noted previously, \emph{DDPM} produces oversmooth noise texture. We further analyze this by isolating a patch of homogeneous soft tissue and computing the power spectral density (PSD) of the noise, as seen in Figure \ref{fig:psd_plot}. One can see that the noise from \emph{DDPM} has an unrealistic concentration of low-frequency components. On the other hand, \emph{DDPM-noise-split} and \emph{DDPM-denoise} have distributed noise characteristics more similar to the input, indicating once again that these schemes are better able to preserve noise texture while still enhancing feature sharpness. Still, these noise spectra are not perfectly representative of the input, and we aim to  future improve noise spectral fidelity.

\section{Conclusion}
In this work, we demonstrate how DDPMs can be combined with simulation methods for PCCT super resolution while better preserving noise texture. Subsequent work will train axial and sagittal plane SR and denoising models, unifying them in the 3D framework presented in \cite{niu_fully3D}. Furthermore, we will improve the noise preservation and transfer learning to the target PCCT data, as well as evaluate the approach on both head and chest images with reader studies.

\section*{Acknowledgment}

Research reported in this publication was supported by GE HealthCare and First Imaging as well as by the National Institutes of Health under Award Numbers R01EB026646, R01CA233888, R01CA237267, R01HL151561, R42GM142394, R21CA264772, R01EB031102, R01EB032716. The content is solely the responsibility of the authors and does not necessarily represent the official views of the National Institutes of Health.

\ifCLASSOPTIONcaptionsoff
  \newpage
\fi


\bibliographystyle{IEEEtran}
\bibliography{IEEEtran/ct_meeting}

\begin{thebibliography}{10}
\providecommand{\url}[1]{#1}
\csname url@samestyle\endcsname
\providecommand{\newblock}{\relax}
\providecommand{\bibinfo}[2]{#2}
\providecommand{\BIBentrySTDinterwordspacing}{\spaceskip=0pt\relax}
\providecommand{\BIBentryALTinterwordstretchfactor}{4}
\providecommand{\BIBentryALTinterwordspacing}{\spaceskip=\fontdimen2\font plus
\BIBentryALTinterwordstretchfactor\fontdimen3\font minus \fontdimen4\font\relax}
\providecommand{\BIBforeignlanguage}[2]{{%
\expandafter\ifx\csname l@#1\endcsname\relax
\typeout{** WARNING: IEEEtran.bst: No hyphenation pattern has been}%
\typeout{** loaded for the language `#1'. Using the pattern for}%
\typeout{** the default language instead.}%
\else
\language=\csname l@#1\endcsname
\fi
#2}}
\providecommand{\BIBdecl}{\relax}
\BIBdecl

\bibitem{willemink_photon-counting_2018}
M.~J. Willemink, M.~Persson, A.~Pourmorteza, N.~J. Pelc, and D.~Fleischmann, ``\BIBforeignlanguage{eng}{Photon-counting {CT}: {Technical} {Principles} and {Clinical} {Prospects}},'' \emph{\BIBforeignlanguage{eng}{Radiology}}, vol. 289, no.~2, pp. 293--312, Nov. 2018.

\bibitem{lim2017enhanced}
B.~Lim, S.~Son, H.~Kim, S.~Nah, and K.~Mu~Lee, ``Enhanced deep residual networks for single image super-resolution,'' in \emph{CVPR}, 2017, pp. 136--144.

\bibitem{zhang2018image}
Y.~Zhang, K.~Li, K.~Li, L.~Wang, B.~Zhong, and Y.~Fu, ``Image super-resolution using very deep residual channel attention networks,'' in \emph{ECCV}, 2018, pp. 286--301.

\bibitem{zhang2019deep}
K.~Zhang, W.~Zuo, and L.~Zhang, ``Deep plug-and-play super-resolution for arbitrary blur kernels,'' in \emph{CVPR}, 2019, pp. 1671--1681.

\bibitem{ledig2017photo}
C.~Ledig, L.~Theis, F.~Husz{\'a}r, J.~Caballero, A.~Cunningham, A.~Acosta, A.~Aitken, A.~Tejani, J.~Totz, Z.~Wang \emph{et~al.}, ``Photo-realistic single image super-resolution using a generative adversarial network,'' in \emph{CVPR}, 2017, pp. 4681--4690.

\bibitem{menon2020pulse}
S.~Menon, A.~Damian, S.~Hu, N.~Ravi, and C.~Rudin, ``Pulse: Self-supervised photo upsampling via latent space exploration of generative models,'' in \emph{CVPR}, 2020, pp. 2437--2445.

\bibitem{ho2020denoising}
J.~Ho, A.~Jain, and P.~Abbeel, ``Denoising diffusion probabilistic models,'' \emph{NeurIPS}, vol.~33, pp. 6840--6851, 2020.

\bibitem{song2022solving}
Y.~Song, L.~Shen, L.~Xing, and S.~Ermon, ``Solving inverse problems in medical imaging with score-based generative models,'' in \emph{ICLR}, 2022.

\bibitem{chung2022improving}
H.~Chung, B.~Sim, D.~Ryu, and J.~C. Ye, ``Improving diffusion models for inverse problems using manifold constraints,'' in \emph{NeurIPS}.

\bibitem{xia2022patch}
W.~Xia, W.~Cong, and G.~Wang, ``Patch-based denoising diffusion probabilistic model for sparse-view {CT} reconstruction,'' \emph{arXiv preprint arXiv:2211.10388}, 2022.

\bibitem{dhariwal2021diffusion}
P.~Dhariwal and A.~Nichol, ``Diffusion models beat {GANs} on image synthesis,'' 2021.

\bibitem{saharia2022image}
C.~Saharia, J.~Ho, W.~Chan, T.~Salimans, D.~J. Fleet, and M.~Norouzi, ``Image super-resolution via iterative refinement,'' \emph{IEEE TPAMI}, 2022.

\bibitem{yu2017computed}
H.~Yu, D.~Liu, H.~Shi, H.~Yu, Z.~Wang, X.~Wang, B.~Cross, M.~Bramler, and T.~S. Huang, ``Computed tomography super-resolution using convolutional neural networks,'' in \emph{ICIP}, 2017, pp. 3944--3948.

\bibitem{you2019ct}
C.~You, G.~Li, Y.~Zhang, X.~Zhang, H.~Shan, M.~Li, S.~Ju, Z.~Zhao, Z.~Zhang, W.~Cong \emph{et~al.}, ``{CT} super-resolution {GAN} constrained by the identical, residual, and cycle learning ensemble ({GAN-CIRCLE}),'' \emph{IEEE TMI}, vol.~39, no.~1, pp. 188--203, 2019.

\bibitem{jiang2020ct}
X.~Jiang, Y.~Xu, P.~Wei, and Z.~Zhou, ``{CT} image super resolution based on improved srgan,'' in \emph{ICCCS}.\hskip 1em plus 0.5em minus 0.4em\relax IEEE, 2020, pp. 363--367.

\bibitem{li2020x}
M.~Li, D.~S. Rundle, and G.~Wang, ``X-ray photon-counting data correction through deep learning,'' \emph{arXiv preprint arXiv:2007.03119}, 2020.

\bibitem{niu_fully3D}
C.~Niu, C.~Wiedeman, M.~Li, J.~Maltz, and G.~Wang, ``{3D} photon counting {CT} image super-resolution using conditional diffusion model,'' in \emph{17th International Meeting on Fully 3D Image Reconstruction in Radiology and Nuclear Medicine}, Stony Brook, NY, USA, 2023.

\bibitem{deman2007catsim}
B.~D. Man, S.~Basu, N.~Chandra, B.~Dunham, P.~Edic, M.~Iatrou, S.~McOlash, P.~Sainath, C.~Shaughnessy, B.~Tower, and E.~Williams, ``{CatSim: a new computer assisted tomography simulation environment},'' in \emph{Medical Imaging 2007: Physics of Medical Imaging}, vol. 6510, 2007, p. 65102G.

\bibitem{wu2022xcist}
M.~Wu, P.~FitzGerald, J.~Zhang, W.~P. Segars, H.~Yu, Y.~Xu, and B.~D. Man, ``{XCIST}—an open access x-ray/{CT} simulation toolkit,'' \emph{Physics in Medicine \& Biology}, vol.~67, no.~19, p. 194002, 2022.

\bibitem{niu_noise2sim_2020}
C.~Niu, M.~Li, F.~Fan, W.~Wu, X.~Guo, Q.~Lyu, and G.~Wang, ``Noise suppression with similarity-based self-supervised deep learning,'' \emph{IEEE Transactions on Medical Imaging}, vol.~42, no.~6, pp. 1590--1602, 2023.

\end{thebibliography}

\end{document}